\documentclass{jps-cp}
\usepackage{txfonts} %Please comment out this line unless the txfonts package is availabe in your LaTeX system.

\title{Understanding recent observations of isolated magnetars and accreting magnetars from the magnetospheric point of view}

\author{H. \textsc{Tong}$^{1,2}$ }

\inst{$^{1}$School of Physics and Electronic Engineering, Guangzhou University, 510006 Guangzhou, China\\
$^{2}$Xinjiang Astronomical Observatory, Chinese Academy of Science, Urumqi, China}

\email{htong\_2005@163.com}

\recdate{May XX, 2017}

\abst{The wind braking model and its applications of magnetars are discussed. The decreasing torque of magnetars during outbursts, anti-glitch, and anti-correlations between radiation and timing are understandable in the wind braking model. Recent 
timing observations of magnetars are also consistent with the previous modeling. A magnetism-powered wind nebula and a braking index smaller than three are the two predictions. Besides isolated magnetars, there may also be accreting magnetars in binary systems and magnetars accreting from fallback disks. Observationally, ultra-luminous X-ray pulsars may be accreting magnetars, while super-slow magnetars may be magnetars with fallback disks in the past. Many works are needed for both isolated magnetars and accreting magnetars.}

\kword{accretion, magnetar, neutron star, pulsar, }

\begin{document}
\maketitle

\section{Introduction}

\begin{figure}[tbh]
\centering
\includegraphics{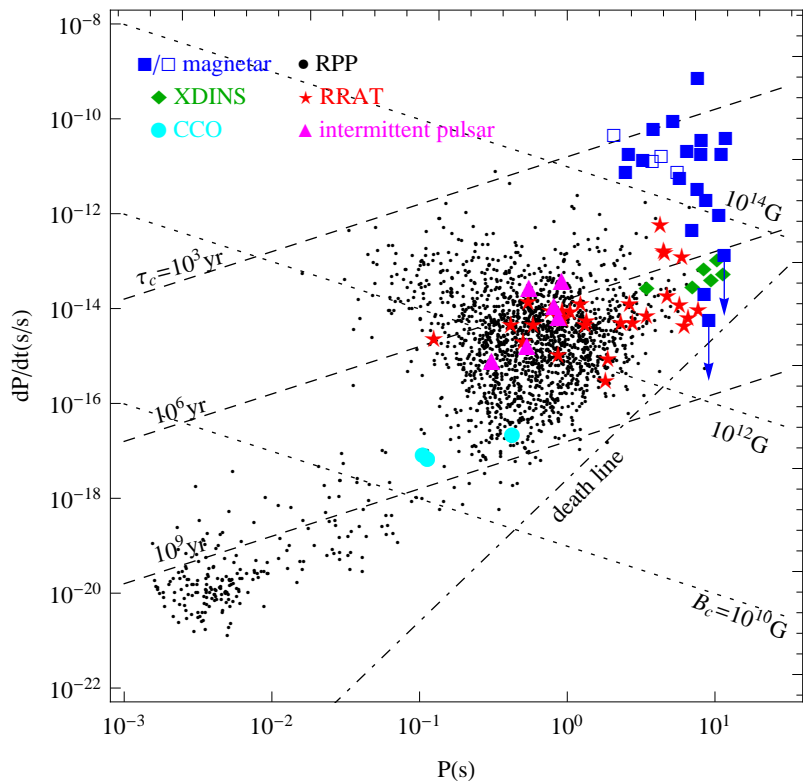}
\caption{Period period-derivative diagram of pulsars. Black points are normal pulsars, blue squares are for magnetars, 
empty blue squares are for radio loud magnetars, updated from figure 1 in \cite{TongWang2014}.}
\label{figure_PPdot}
\end{figure}

After 50 years of their discovery, pulsars are still fascinating objects. They possess the strongest magnetic fields 
in the universe. Magnetars is a special kind of pulsars. They may have the strongest magnetic field in all pulsars, with magnetic field as high as $10^{15} \,\rm G$ \cite{Mereghetti2008,Mereghetti2015}. Therefore, magnetars are new specimen to study the physics of normal pulsars and accretion-powered X-ray pulsars \cite{Tong2015a,Tong2016a}. For isolated magnetars, their distribution on the period period-derivative diagram of pulsars is shown in figure \ref{figure_PPdot}. As can be seen this figure, both pulsars and magnetars are spinning down. One fundamental question of pulsar study is: what's the braking mechanism of pulsars and magnetars? In both pulsar and magnetar studies, the magnetic dipole braking mechanism is usually assumed. However, this is only a crude approximation since it assumes a rotating dipole in vacuum. 
A physical braking mechanism must consider the presence of the pulsar magnetosphere. There are many models in this direction. Among them are the wind braking of pulsars \cite{KouTong2015} and wind braking of magnetars \cite{Tong2013}.

Observationally, magnetars have varying spin-down torque \cite{Archibald2015a}, or decreasing torque during outburst \cite{Scholz2017}, or increasing torque \cite{Younes2015}. Their spectra also contain a significant fraction of nonthermal emissions \cite{Weng2015,Enoto2017}. Therefore, a physical picture is that: the magnetic energy of magnetars is first converted to a system of particles. These particles are responsible for both the X-ray emissions and the spin-down torque. Then, it is natural that correlations between radiation and timing are seen ubiquitously in magnetars. This means that the spin-down of magnetars may be  dominated by the outflowing particles. This is dubbed as the ``wind braking of magnetars'' \cite{Tong2013}. 

Recently, there are many interesting observations of magnetars. It is found that many of them can be understood safely in the wind braking model. Furthermore, the discovery of possible accreting magnetars enable people to study the magnetosphere of accreting magnetars. In the following, the recent observations of isolated magnetars and accreting magnetars are discussed.  Understandings of these observations from the magnetospheric point of view are presented.  

\section{Isolated magnetars: wind braking and wind nebula}

\subsection{Decreasing torque during outbursts}

In the wind braking scenario, magnetars are neutron stars with strong multipole field. The strong multipole field is responsible for
the X-ray emission, bursts, and spin-down of the central magnetar. The rotation energy of the magnetar is carried away mainly the outflowing particles. Their surface dipole field can be very high or not very high. The discovery of low magnetic field magnetars is consistent with the picture of wind braking of magnetars. For the second low magnetic field magnetar Swift J1822.3$-$1606, different authors gave different timing solutions. Ref. \cite{TongXu2013} applied the wind braking model to Swift J1822.3$-$1606. If the particle outflow decreases as that of the X-ray luminosity, then a decreasing period derivative is expected. This may solve the observational ambiguities. Ref. \cite{TongXu2013} predicted a long term period derivative of $1.9\times 10^{-14}$ (last paragraph in section 2 there). 
Subsequent timing found a long term period derivate of $(2.1\pm 0.2) \times 10^{-14}$ \cite{Scholz2014}. This is consistent with the theoretical prediction. Alternative explanation is discussed in Ref. \cite{Scholz2014}. 

Recent observations of the first radio-loud magnetar XTE J1810$-$197 found a decreasing spin-down torque, accompanied by a decreasing X-ray luminosity and radio luminosity. The radio luminosity finally disappears \cite{Camilo2016}. The spin-down torque of XTE J1810$-$197 decreased by a factor of three during the process. While its radio luminosity decrease by factor of ten \cite{Camilo2016}. Similar radiation and timing behavior is also seen in the third radio-loud magnetar PSR J1622$-$4950 \cite{Scholz2017}.  For PSR J1622$-$4950, its spin-down decrease by about a factor of 10. During this process, the radio flux decrease by a factor about 100 \cite{Scholz2017}. In the wind braking scenario, the spin-down rate is proportional to the square-root of the particle luminosity: $\dot{P} \propto L_{\rm p}^{1/2}$ \cite{Tong2013}. These particles may be also responsible for the radio and X-ray emission. One possibility is that a constant fraction of the particle luminosity is converted to X-ray and radio emission. Then the decrease in radio luminosity and decrease in spin-down torque for these two radio emitting magnetars are quantitatively consistent with each other. X-ray observations of XTE J1810$-$197 found a decrease in spin-down torque by a factor of eight \cite{Camilo2016} during the whole outburst epoch. While the X-ray flux decreased by a factor of 50 \cite{Alford2016}. This is also consistent with the anticipation in the wind braking model. 

Furthermore, in the wind braking scenario, the magnetars is expect to have smaller dipole magnetic field than their apparent characteristic magnetic field. Then during quiescent state, the magnetar may be below the radio death line of pulsars, i.e. they are radio quiet. During outburst, when the magnetosphere of magnetars is perturbed significantly, the magnetar may be radio-loud temporally. After the outburst, when the magnetar return back to the quiescent state, it will become radio-quiet again (Ref. \cite{Lin2015} obtained a similar conclusion using a different model for magnetars). The final radio disappearance of these two radio-loud magnetars \cite{Camilo2016,Scholz2017} are consistent the general expectations in the wind braking model. 

\subsection{Anti-glitch, anti-correlation between radiation and timing}

Glitches in normal pulsars and magnetars are sudden spin-up events \cite{Dib2014}. However, a spin-down glitch (i.e. anti-glitch) is observed in one magnetar 1E 2259$+$586 \cite{Archibald2013}. Meanwhile, the X-ray luminosity also increases by about a factor of two during this epoch. The X-ray luminosity contains a significant non-thermal component. The increase in X-ray luminosity should be accompanied by an increase in the particle luminosity. This will result in an enhanced rotational energy loss rate. It is found that the rotational energy carried away by the outflowing particles may explain the  so-called ``anti-glitch'' \cite{Tong2014}. In this case, there is no ``anti-glitch''. It is is just a period of enhanced spin-down caused by an enhanced particle wind. In the original work for anti-glitch \cite{Tong2014}, the net spin-down glitch of PSR J1846$-$0258 is also discussed. Later observations of spin-down glitches in magnetar 4U 0142$+$61 etc are consistent with the wind braking model \cite{Archibald2017}. 

Timing of the Galactic center magnetar SGR J1745$-$2900 found a decreasing X-ray luminosity accompanied by an increasing spin-down torque \cite{Kaspi2014}. For a constant particle luminosity, a change in the polar cap geometry may explain the anti-correlation between X-ray luminosity and spin-down torque \cite{Tong2015b}. Subsequent timing of SGR 1806$-$20 also found an increasing spin-down torque \cite{Younes2015}. It is qualitatively consistent with the change in emission geometry. Detailed modeling is needed for this source. 

In summary, there are two typical changes in the polar caps of magnetar magnetospheres: (1) The total particle luminosity changes while the geometry is unchanged. The luminosity can increase or decrease, while the pulse profile is almost constant. This will result in a positive correlation between radiation and spin-down torque. This may correspond to the decreasing spin-down torque during the outburst of magnetars. And an enhanced spin-down period may result in a net spin-down of the magnetar. 
(2) The geometry of the polar cap changes while the particle luminosity is almost constant. This will result in a anti-correlation between the radiation and timing of the magnetar. Changes in the pulse profile is also possible. In reality, both the particle luminosity and geometry may change, this may corresponding to the diverse behaviors of magnetars radiation and timing. 

\subsection{Two predictions: wind nebula and braking index}

There are two predictions of the wind braking model: a magnetism-powered wind nebula and a braking index smaller than three \cite{Tong2013}. The particle luminosity of magnetars may be comparable with their X-ray luminosities, about $10^{35} \rm \, erg \, s^{-1}$. It is usually much higher than the magnetar's rotational energy loss rate. However, it is lower than  the rotational energy loss rate of young pulsars, which is from about $10^{36} \rm\, erg \,s^{-1}$ to $10^{38} \rm\, erg \, s^{-1}$. Pulsar wind nebulae are always seen around young pulsars. Therefore, the particle outflow of magnetars may also be seen as a wind nebula around the magnetar. Since the particle outflow originates from the magnetic energy release, the corresponding wind nebula is powered by the magnetic energy rather than the magnetar's rotational energy. Because the particle luminosity is lower than that of the young pulsars', the magnetar wind nebula may also be relatively weak. This may explain why magnetar wind nebula is not detected until 2016 \cite{Younes2016}. The possible wind nebula around magnetar Swift J1834.9$-$0846 \cite{Younes2016} may be a magnetism-powered wind nebula \cite{Tong2016b}. The structure and evolution of magnetar wind nebula need further studies. 

For both normal pulsars and magnetars, a particle wind will result in a braking index about one. Observationally magnetars are more noisy \cite{Dib2014}. Therefore, the braking index of magnetars may be difficult to measure. However, there are some indirect evidences. The rotation-powered pulsar PSR J1846$-$0258 showed some kind of magnetar-like activities. Later timing found a smaller braking index \cite{Archibald2015b,Kou2016}. The radio pulsar PSR J1119$-$6127 have braking index measurement. Recently, it also showed some kind of magnetar-like activities \cite{Archibald2016}. One anticipation of the wind braking model is that: PSR J1119$-$6127 will have a smaller braking index after the outburst. This may be tested by future observations.  

\section{Accreting magnetars}

Magnetar is a special kind of pulsars. Since there are both isolated pulsars and accretion-powered X-ray pulsars in binary systems, then both isolated magnetars and accreting magnetars should exist. For accreting magnetars, the question is: which accreting neutron star is an accreting magnetar? In order to answer this question and finding accreting magnetars, we must find the difference between accreting normal neutron stars and accreting magnetars. By analyzing previous studies of pulsars and magnetars, we proposed that accreting magnetars can have two signatures \cite{TongWang2014}: (1) a magnetar-like outburst, (2) a hard X-ray tail above $100\,\rm keV$. Ultraluminous X-ray pulsars are accretion-powered X-ray pulsars with luminosity as high as or higher than $10^{40} \,\rm erg \,s^{-1}$ \cite{Bachetti2014}. With the discovery of ultraluminous X-ray pulsars, it is proposed that they may also be accreting magnetars \cite{Tong2015c}. Considering that the magnetar in the binary system may be an old magnetar. Old magnetar are more likely to be low magnetic field magnetars. In the frame of accreting low magnetic field magnetar, both the radiation and timing behaviors of ultraluminous X-ray pulsar can be explained. Up to now, three ultraluminous X-ray pulsars are discovered \cite{Israel2017}. The radiation and evolution of accreting magnetars need more studies. 

Neutron stars are born in supernova explosions. During this process, some of the ejecta may fallback and form a disk, i.e. a fallback disk.
Neutron stars accreting from a fallback disk is proposed as an alternative model to the magnetar model. Some evidence for a fallback disk is also found. Since magnetars are also born in supernova explosions, then they can also have fallback disks. 
The high magnetic field of magnetars, combined with a relatively low accretion rate from the fallback disk, will result in a long equilibrium period. The central compact object inside supernova remnant RCW 103 is identified as a magnetar with a possible rotational period of $2\times 10^4 \,\rm s$ \cite{DAi2016,Rea2016}. This super-slow magnetars may have a fallback disk in the past \cite{Tong2016c}. This can account for its super-slow rotation. Now the disk has faded away, and the central neutron star can have magnetar activities. Pulsars and magnetars with fallback disks may explain a wide range of peculiar observations. 
 
\section{Summary}

By studying recent observations of magnetars, we may get the following conclusions:
\begin{enumerate}
\item Anomalous X-ray pulsars and soft gamma-ray repeaters may be magnetars. The spin-down of magnetars may be dominated by the particle wind. In the wind braking model of magnetars, it is natural that there are correlations between the radiation and timing of magnetars. 

\item Magnetars may also generate a wind nebula. This wind nebula can be powered by the magnetic energy of magnetars. 

\item Central compact objects may be magnetars in waiting or magnetars with fallback disks. 

\item Ultraluminous X-ray pulsars may be another manifestation of accreting magnetars. 

\end{enumerate}

In summary, magnetar is just a special kind of pulsars. It is special because it can have a higher magnetic field and can be powered by the magnetic energy. At the same time, magnetar can also have wind nebula, fallback disk, and accretion from a binary companion, similar to that of normal pulsars.  

\section*{Acknowledgement}

H. Tong would to thank his collaborators very much, including R. X. Xu, W. Wang, F. F. Kou etc, and C. H. Lee etc for discussions.

\end{document}